\documentclass[12pt]{article}
\pdfoutput=1

\usepackage{a4}
\usepackage{a4wide}

\usepackage{amsmath}
\usepackage{amscd}
\usepackage{amsfonts}
\usepackage{amssymb}
\usepackage{mathrsfs}
\usepackage{fancybox} 
\usepackage{enumerate}
\usepackage{bm}
\usepackage{amscd}
\usepackage{graphicx}
\usepackage{epsfig}

\makeatletter

\@addtoreset{equation}{section}
\makeatother
\usepackage{color}
\definecolor{blue1}{rgb}{0.15,0.15,0.50}
\PassOptionsToPackage{hyphens}{url}
\usepackage{hyperref}
\usepackage{amsthm} % This is for \newtheorem*

\newtheorem*{pf}{Proof}
\def\={\stackrel{\bullet}{=}}

\def\({\left(}
\def\){\right)}
\def\[{\left[}
\def\]{\right]}

\def \be {\begin{equation}}
\def \ee {\end{equation}}
\def \beal#1 {\begin{align}#1\end{align}}
\def \bes#1 {\begin{equation}\begin{split}#1\end{split}\end{equation}}

\makeatletter

\@addtoreset{equation}{section}
\makeatother

\begin{document}

\begin{titlepage}

\title{
\vspace{-2cm}
\begin{flushright}
\normalsize{UT-Komaba/20-2
\\
YITP-20-63}
\end{flushright}
       \vspace{1.5cm}
       {\LARGE  $U(1)$ spin Chern-Simons theory \\
         and Arf invariants in two dimensions}       
\vspace{1.5cm}
}

\author{
Takuya Okuda\thanks{takuya[at]hep1.c.u-tokyo.ac.jp},\; Koichi Saito\thanks{saito[at]hep1.c.u-tokyo.ac.jp},\; Shuichi Yokoyama\thanks{shuichi.yokoyama[at]yukawa.kyoto-u.ac.jp
}
\\[25pt] 
${}^*{}^{\dagger}${\it Graduate School of Arts and Sciences, University of Tokyo}\\
{\it Komaba, Meguro-ku, Tokyo 153-8902, Japan}
\\[10pt]
${}^{\ddagger}${\it Center for Gravitational Physics,} \\
{\it Yukawa Institute for Theoretical Physics, Kyoto University,}\\
{\it Kitashirakawa-Oiwakecho, Sakyo-Ku, Kyoto, Japan}
\\[10pt]
}

\date{}
\maketitle

\thispagestyle{empty}

% \vspace{.2cm}

\begin{abstract}
\vspace{0.3cm}
\normalsize

\noindent
The level-$k$ $U(1)$ Chern-Simons theory is a spin topological quantum field theory for $k$ odd.
Its dynamics is captured by the 2d CFT of a compact boson with a certain radius.
Recently it was recognized that a dependence on the 2d spin structure can be given to the CFT by modifying it using the so-called Arf invariant.
We demonstrate that one can reorganize the torus partition function of the modified CFT into a finite sum involving a finite number of conformal blocks.
This allows us to reproduce the modular matrices of the spin theory.
We use the modular matrices to calculate the partition function of the spin Chern-Simons theory on the lens space~$L(a,\pm 1)$, and demonstrate the expected dependence on the 3d spin structure.

\end{abstract}

\end{titlepage}

\tableofcontents

%%%%%%%%%%%%%%%%%%%%%%%%%%%%%%%%%%%%%
\section{Introduction} 
\label{sec:intro}

In recent years there has been much progress in understanding the topological phases of matter.
Particularly interesting is the case where the underlying physical system involves fermions.
At low energies such a system realizes a fermionic topological phase, described by a topological quantum field theory (TQFT) sensitive to the spin structure of spacetime.
The aim of this note is to study a simple example of spin TQFT by extending elementary techniques~\cite{DiFrancesco:1997nk} familiar from the study of 2d conformal field theories (CFTs).
While we focus on the most basic CFT, we believe that our results and considerations will be useful in broader contexts.

It is well-known that the CFT of a free compact boson $\phi$ of radius $R$ is rational when $R^2$ is rational.
(We use the normalization such that T-duality acts as $R\rightarrow 2/R$.)
For simplicity let $k$ be a positive integer and set $R=2/\sqrt{k}$ (or its T-dual $R=\sqrt{k}$).
By the bulk-boundary correspondence, this ``edge state'' CFT characterizes the dynamics of the $U(1)$ level-$k$ Chern-Simons (CS) theory~\cite{Moore:1989yh}.%
\footnote{\label{footnote:glue-chiral-CFTs}%
More precisely the full CFT that glues the left- and right-moving chiral CFTs corresponds to the 3d theory on an interval~$[0,2]$ times a 2d space~$\Sigma$.
The chiral CFTs live on~$\{0\}\times\Sigma$ and $\{2\}\times\Sigma$.
}
We consider the extended chiral algebra generated by
\begin{equation}\label{generators}
e^{\pm i \sqrt{k} \phi_L}
\end{equation}
in addition to $i\partial\phi_L$, where~$\phi_L$ is the left-moving part of $\phi$.
See Appendix C.1 of~\cite{Seiberg:2016rsg} for a modern discussion.
For $k$ even, the chiral operators~(\ref{generators}) are in the physical spectrum, and
 the usual torus partition function can be written as a finite sum in terms of a finite number of characters of the algebra~\cite{Moore:1988ss}.
 (See Appendix~\ref{app:even}.)

The case of~$k$ odd is somewhat more subtle.
As we will see, the would-be generators~(\ref{generators}) of an extended chiral algebra are not in the physical spectrum of the boson theory.%
\footnote{\label{footnote-U14k}%
The chiral operators $e^{\pm 2i \sqrt{k} \phi_L}$ corresponding to $U(1)_{4k}$ are in the physical spectrum, and the boson torus partition function can be written as a finite sum in terms of a finite number of characters for the algebra they generate.
The~$U(1)_{4k}$ CS theory is obtained from the~$U(1)_{k}$ CS theory by gauging the fermionic parity, {\it i.e.}, by summing over spin structures~\cite{Gaiotto:2015zta,Bhardwaj:2016clt}.
The former is called the ``shadow'' of the latter, and they are related via a process called ``fermionic anyon condensation.''
}
Also, the $U(1)_k$ Chern-Simons theory is a spin topological quantum field theory~\cite{Dijkgraaf:1989pz} although the CFT is bosonic and does not depend on the 2d spin structure.
We will see that these issues are nicely resolved by modifying the bosonic CFT into a spin CFT according to the recently proposed procedure involving the so-called Arf invariant~\cite{Kapustin:2017jrc,YTCernLect,Karch:2019lnn}.%
\footnote{For the earlier related literature see the references mentioned in~\cite{Harvey:2020jvu}.
The procedure was recently applied to minimal models in~\cite{Hsieh:2020uwb,Kulp:2020iet} to obtain new types of fermionic minimal models~\cite{Runkel:2020zgg}.}

In this note we demonstrate that one can rewrite the torus partition function of the modified theory as a finite sum in terms of a finite number of spin structure dependent conformal blocks of the extended chiral algebra.
(For $k=1$ and $R=2$, this was done in~\cite{Karch:2019lnn}.)
We use the conformal blocks to compute the modular matrices of the modified compact boson CFT.
They coincide with the matrices obtained from the Chern-Simons theory~\cite{Belov:2005ze,Stirling:2008bq} up to conjugation.
We then use the modular matrices to compute the partition function of the $U(1)_k$ spin Chern-Simons theory on the lens space $L(a,\pm 1)$.
For $k$ odd and $a$ even, we obtain the expected dependence on the spin structure on $L(a,\pm 1)$.

This paper is organized as follows.
In Section~\ref{sec:CFT} we study the free boson theory modified by the Arf invariant.
We first review the modification procedure as described in~\cite{Karch:2019lnn}.
We then expand the modified torus partition function in terms of a finite number of conformal blocks.
Using the conformal blocks we compute the modular matrices.
In Section~\ref{sec:CS} we use the modular matrices to compute the partition functions on~$L(a,\pm1)$.
In Appendix~\ref{app:even}, we summarize the modular matrices and the $L(a,\pm1)$ partition functions for $k$ even.
In Appendix~\ref{app:quad-Arf}, we review the notions of quadratic refinements and their Arf invariants.

\section{Modular matrices of the compact boson spin CFT}\label{sec:CFT}

We consider the theory $\mathcal{T}_\phi$ of a free boson $\phi$ parametrizing the circle of radius $R$ ($\phi\sim \phi+2\pi R$) with an action
\begin{equation}
S= \frac{1}{8\pi} \int d^2 x\, \partial_\mu  \phi \, \partial^\mu \phi \,.
\end{equation}
In this normalization T-duality acts as $R\rightarrow 2/R$.
Let $\tau$ be the modulus of the torus and set $q=e^{2\pi i \tau}$.
The torus partition function is given as the sum over the physical spectrum
\begin{equation}\label{torus-pf-phi}
  Z[\mathcal{T}_\phi]
=
\frac{1}{|\eta(\tau)|^2 }
 \sum_{n, w\in\mathbb{Z}} 
  q^{\frac{1}{2}(\frac{n}{R} + \frac{w R}{2})^2} \bar q^{\frac{1}{2}(\frac{n}{R}-\frac{w R}{2})^2} \,,
\end{equation}
where $\eta(\tau)$ is the Dedekind eta function.
Each term corresponds to the local operator
\begin{equation}
 e^{i p_L \phi_L + i p_R \phi_R} 
\end{equation}
with $ p_L = \frac{n}{R} + \frac{w R}{2}$, $   p_R = \frac{n}{R} - \frac{w R}{2} $ for $n,w\in\mathbb Z$.
For generic $R$ the chiral algebra is generated by $i\partial \phi_L$.

We are interested in the so-called rational boson, for which the radius-squared is a rational number.
We set $R=2 (p/p'')^{1/2}$ with $p$ and $p''$ relatively prime positive integers.
We also set%
\footnote{%
The $R=2/\sqrt{k}$ case mentioned in the introduction corresponds to $p=1$, $p''=k$.
According to~\cite{Kapustin:2010if}, where the case with $k$ even was studied, the integers~$(p,p'')$ specify the type of a domain wall placed on~$\{1\}\times\Sigma$ in the set-up of footnote~\ref{footnote:glue-chiral-CFTs}.
}
\begin{equation}
k:=pp'' \,.
\end{equation}
The chiral operators~$e^{\pm i\sqrt k \phi_L}$ with $(p_L,p_R)=(\pm \sqrt k, 0)$ correspond to $(n,w)=\pm (p,p''/2)$.
For $p''$ even, hence with $k$ even, these are part of the physical spectrum, and extend the chiral algebra.
For $p''$ odd, and especially when $k$ is odd, however, they are not in the spectrum.

\subsection{Modification by the Arf invariant}\label{sec:Arf}

We now introduce the spin structure dependence into the 2d CFT following~\cite{Karch:2019lnn}.

The crucial ingredient is the topological theory that we call~$\mathcal{T}_\text{Arf}$.
It is the low-energy limit of the Kitaev Majorana chain~\cite{Kitaev:2001kla} and has the partition function given as
\begin{equation}\label{TArf-PF}
Z[\mathcal{T}_\text{Arf};\rho]=e^{\pi i \text{Arf}[\rho]} \,,
\end{equation}
where~$\text{Arf}[\rho]$ is the so-called Arf invariant%
\footnote{%
More precisely, this is the Arf invariant of a quadratic refinement of the intersection pairing on~$H_1(\Sigma,\mathbb Z_2)$, which is in one-to-one correspondence with a spin structure on the surface~$\Sigma$~\cite{MR588283}.
The Arf invariant also appears in~\cite{Belov:2005ze}, where it is called the mod 2 index, but it plays different roles.
} 
determined by the spin structure $\rho$ on a 2d surface.
It is identified with the index mod 2 of the Dirac operator given by the spin structure~$\rho$.
In particular we have
\begin{equation}
\text{Arf}[\rho] =
\left\{
\begin{array}{ll}
1 & \text{ if } \rho=PP \,, \\
0 &  \text{ if }  \rho=PA\,, AP\,, AA 
\end{array}
\right.
\end{equation}
for the four spin structures on the torus, where $P$ and $A$ denote periodic and anti-periodic boundary conditions, respectively.%
\footnote{%
For example, $PA$ corresponds to the periodic boundary condition in the space (horizontal) direction and the anti-periodic boundary condition in the time (vertical) direction.
}

On a general surface~$\Sigma$, a $\mathbb{Z}_2$ gauge field $S$ is an element of $H^1(\Sigma,\mathbb{Z}_2)$ and acts on the spin structure: $\rho \rightarrow S\cdot \rho$.
Indeed  a $\mathbb{Z}_2$ gauge field can be regarded as a $\mathbb{Z}_2$ holonomy and it modifies the boundary condition of a fictitious spinor along a closed path dictated by~$\rho$.
The holonomy along the horizontal (resp. vertical) direction can be regarded as the insertion of a topological defect along the vertical (resp. horizontal) direction.
The defect is the symmetry generator of the 0-form ({\it i.e.}, ordinary) $\mathbb{Z}_2$ global symmetry~\cite{Gaiotto:2014kfa}.
Thus the topological theory $\mathcal{T}_\text{Arf}$ has a global $\mathbb{Z}_2$ symmetry, even though it has no local operators on which the symmetry acts.

The boson theory~$\mathcal{T}_\phi$ also has a symmetry $\mathbb{Z}_2^S$ generated by the shift $\phi \rightarrow \phi +\pi R$.%
\footnote{%
%This symmetry is denoted as $\mathbb{Z}_2^S $ in~\cite{Karch:2019lnn}.
Another symmetry~$\mathbb{Z}_2^C$ generated by $\phi \rightarrow -\phi$ would replace~$\mathbb{Z}_2^S$ if we work in the T-dual frame.
}
We now consider the theory 
\begin{equation}
(\mathcal{T}_\phi\times \mathcal{T}_\text{Arf})/\mathbb{Z}_2^\text{diag} \,,
\end{equation}
obtained by gauging
 the diagonal subgroup~$\mathbb{Z}_2^\text{diag}$ of the product of $\mathbb{Z}_2^S$ and the symmetry group~$\mathbb{Z}_2$ of~$\mathcal{T}_\text{Arf}$.
To write down formulas for the torus partition function succinctly, we choose a reference spin structure on the torus, say $\rho_0=AA$,%
\footnote{%
This choice is motivated by the labeling (characteristics) of the theta functions~$\vartheta_{ab}(\nu;\tau)$.
See~(\ref{L-theta}).
} 
and identify general spin structures $\rho$ with $\mathbb{Z}_2$ gauge fields~$S=(S_1,S_2)$ via $\rho=S\cdot\rho_0$:
\begin{equation}
AA \leftrightarrow S=(0,0) \,, \quad
AP \leftrightarrow (0,1) \,, \quad
PA \leftrightarrow (1,0) \,, \quad
PP \leftrightarrow (1,1) \,.
\end{equation}
To compute the torus partition function 
we sum over dynamical $\mathbb{Z}_2^\text{diag}$ gauge fields $s=(s_1,s_2)$ and divide by $|\mathbb{Z}_2^\text{diag}|=
2$.
We get%
\footnote{%
This is equivalent to (3.28) of~\cite{Karch:2019lnn} with $S_\text{there}=C=0$.
}
\begin{equation}\label{Z-phi-Arf-S}
\begin{aligned}
&  Z[(\mathcal{T}_\phi\times \mathcal{T}_\text{Arf})/\mathbb{Z}_2^\text{diag};R ; \rho=S\cdot \rho_0]
\\
& \qquad = 
\frac{1}{2|\eta(\tau)|^2 } \sum_{s=(s_1,s_2)} 
 (-1)^{(S_1+s_1)(S_2 + s_2)}
 \sum_{n\in\mathbb{Z}} \sum_{w\in \mathbb{Z}+\frac{1}{2}s_1}
 (-1)^{n s_2}
  q^{\frac{1}{2}(\frac{n}{R} + \frac{w R}{2})^2} \bar q^{\frac{1}{2}(\frac{n}{R}-\frac{w R}{2})^2} \,.
\end{aligned}
\end{equation}
The sign~$(-1)^{(S_1+s_1)(S_2 + s_2)}$
is the partition function~(\ref{TArf-PF}) of~$\mathcal{T}_\text{Arf}$ for the spin structure~$(S+s)\cdot\rho_0$.
The winding number~$w$ is a half-odd integer in the twisted sector ($s_1=1$).
The momentum~$n$ induces a sign~$(-1)^n$ under the shift $\phi\rightarrow \phi+\pi R$.
Explicitly, we have%
\footnote{%
We have the relations $n=m+\bar m + S_1$, $w=(m-\bar m)/2$.
}
\begin{equation}\label{torus-pf-modified}
\begin{aligned}
& \quad Z[(\mathcal{T}_\phi\times \mathcal{T}_\text{Arf})/\mathbb{Z}_2^\text{diag};R ;\rho]
= 
\frac{1}{|\eta(\tau)|^2} \times
\\
&
\hspace{-2mm}
\times
\hspace{-2mm}
\sum_{m,\bar{m}\in\mathbb{Z}}   \left\{
\begin{array}{lll}
q^{
\frac{1}{8} \left(
m\left( \frac{2}{R} +\frac{R}{2}\right) + \bar{m} \left( \frac{2}{R} -\frac{R}{2}\right)
\right)^2
}
\bar{q}^{
\frac{1}{8} \left(
\bar{m}\left( \frac{2}{R} +\frac{R}{2}\right) + m \left( \frac{2}{R} -\frac{R}{2}\right)
\right)^2
}
&
\rho =AA \,, 
\\
(-1)^{m+\bar{m}}
q^{
\frac{1}{8} \left(
m\left( \frac{2}{R} +\frac{R}{2}\right) + \bar{m} \left( \frac{2}{R} -\frac{R}{2}\right)
\right)^2
}
\bar{q}^{
\frac{1}{8} \left(
\bar{m}\left( \frac{2}{R} +\frac{R}{2}\right) + m \left( \frac{2}{R} -\frac{R}{2}\right)
\right)^2
}
&
\rho =AP \,, 
 \\
 q^{
\frac{1}{8} \left(
m\left( \frac{2}{R} +\frac{R}{2}\right) + \bar{m} \left( \frac{2}{R} -\frac{R}{2} \right) + \frac{2}{R} 
\right)^2
}
\bar{q}^{
\frac{1}{8} \left(
\bar{m}\left( \frac{2}{R} +\frac{R}{2}\right) + m \left( \frac{2}{R} -\frac{R}{2}  \right)+ \frac{2}{R}
\right)^2
}
 &
\rho =PA \,, 
 \\
 (-1)^{m+\bar{m}+1}
q^{
\frac{1}{8} \left(
m\left( \frac{2}{R} +\frac{R}{2}\right) + \bar{m} \left( \frac{2}{R} -\frac{R}{2} \right) + \frac{2}{R} 
\right)^2
}
\bar{q}^{
\frac{1}{8} \left(
\bar{m}\left( \frac{2}{R} +\frac{R}{2}\right) + m \left( \frac{2}{R} -\frac{R}{2}  \right)+ \frac{2}{R}
\right)^2
}
 &
\rho =PP \,.
\end{array}
  \right.
\end{aligned}
\end{equation}
These partition functions should be interpreted as ${\rm Tr}_{\mathcal H_*} q^{L_0-1/24}\bar q^{\bar L_0-1/24}$ for $\rho=*A$ and as ${\rm Tr}_{\mathcal H_*} (-1)^Fq^{L_0-1/24}\bar q^{\bar L_0-1/24}$ for $\rho=*P$, where $*$ is either $A$ or $P$.

In the NS sector ($*=A$ in the spatial direction), the physical spectrum is given by
\begin{equation}
p_L =  \frac12\left[m\left( \frac{2}{R} +\frac{R}{2}\right) + \bar{m} \left( \frac{2}{R} -\frac{R}{2}\right)\right]
\,,\ 
p_R = \frac12\left[\bar m\left( \frac{2}{R} +\frac{R}{2}\right) + m \left( \frac{2}{R} -\frac{R}{2}\right)\right] 
\end{equation}
with $m,\bar m\in\mathbb Z$.
The Fermion number is $(-1)^F=(-1)^{m+\bar m}$.

In the R sector ($*=P$ in the spatial direction), the physical spectrum is given by
\begin{equation}
p_L =\frac12
\left[m \hspace{-.5mm}
\left(
 \frac{2}{R} 
 +\frac{R}{2}\right) + \bar{m} \left( \frac{2}{R} -\frac{R}{2}\right) + \frac 2 R\right] 
 \,,
p_R = \frac12\left[\bar m\left( \frac{2}{R} +\frac{R}{2}\right) + m \left( \frac{2}{R} -\frac{R}{2}\right) + \frac 2 R \right] 
\end{equation}
with $m,\bar m\in\mathbb Z$.
The Fermion number is $(-1)^F=(-1)^{m+\bar m+1}$.

For~$R=2 (p/p'')^{1/2}$ and $k=pp''$, the generators~$e^{\pm i \sqrt k\phi_L}$ of the extended chiral algebra have $(p_L,p_R)=(\pm \sqrt k ,0)$.
For $p$ and $p''$ both odd, they are physical and are in the NS sector, with $(m,\bar m) = (\frac{p+p''}{2}, \frac{p - p''}{2})$.

\subsection{Torus conformal blocks}\label{sec:character}

We can expand the spin structure dependent torus partition function~(\ref{torus-pf-modified}), or its generalization by $U(1)$ symmetries, as a finite sum in terms of a finite number of conformal blocks.%
\footnote{%
Our method is brute force.
This result can be obtained more systematically by applying the orbifold method of~\cite{Hsieh:2020uwb} to the compact boson and the $\mathbb{Z}_2$ symmetry generated by~$\phi \rightarrow \phi +\pi R$.
}

We need some preparation.
We define 
\begin{equation}
p_1:=\frac{p''+p}{2}
\,,\qquad
  p_2:=\frac{p''-p}{2} \,,
\end{equation}
which are relatively prime integers because $p$ and $p''$ are odd and relatively prime.
Let us choose $m_0,\bar{m}_0\in\mathbb{Z}$ such that $ m_0 p_1 + \bar{m}_0 p_2 =1$ and set 
\begin{equation}
\omega:= m_0 p_2 + \bar{m}_0 p_1 \,.
\end{equation}
It can be shown that
\begin{equation}
\omega(m p_1+\bar{m} p_2) = m p_2+\bar{m} p_1 \quad \text{ mod } k \,.
\end{equation}

The compact boson has two $U(1)$ symmetries whose charges are the momentum and the winding number.
Their linear combinations give left- and right-moving $U(1)$ symmetries.
We let $z  := e^{2\pi i \nu}$ and $\bar{z}:=e^{-2\pi i \bar\nu}$ be their corresponding fugacities.

Let us define the function
\begin{equation} \label{AppB-K-def}
K_\lambda(z,\tau):= \frac{1}{\eta(\tau)} \sum_{n\in\mathbb{Z}} q^{\frac{(kn+\lambda)^2}{2k}} z^n \,,
\end{equation}
and the {\it conformal blocks} $L^{(S_1,S_2)}_\lambda(\nu,\tau)$ ($S_{1,2}\in\{0,1\}$)%
\footnote{%
These are essentially the specialization of the conformal blocks (the physical wave functions) considered in~\cite{Belov:2005ze} to the case with genus one and gauge group~$U(1)$.
}
\begin{equation}\label{def:conformal-blocks}
\begin{array}{lllll|}
L^{(S_1,S_2)}_\lambda (\nu,\tau)  &\hspace{-2mm}=&\hspace{-2mm}  i^{S_1 S_2}  z^{\lambda/k} K_\lambda((-1)^{S_2}z,\tau)  \,, \quad \lambda\in\mathbb{Z}  +\frac{S_1}{2} \,.
\end{array}
\end{equation}
The blocks~$L^{(S_1,0)}_\lambda$ ($S_1=0,1$) coincide with the characters of the extended chiral algebra
\begin{equation}
L^{(S_1,0)}_\lambda (\nu,\tau)  = {\rm Tr}_{\mathcal{V}_\lambda} q^{L_0-\frac{1}{24}} z^{J_0} \,,
\end{equation}
where~$J_0=p_L/\sqrt{k}$, and $\mathcal{V}_\lambda$ is the representation that contains the state corresponding to the chiral operator $e^{ i \frac{\lambda}{\sqrt{k}}\phi_L}$.%
\footnote{%
The functions~$L^{(S_1,1)}_\lambda$  ($S_1=0,1$) are roughly the ``supercharacters"  ${\rm Tr}_{\mathcal{V}_\lambda} (-1)^{F_L} q^{L_0-\frac{1}{24}} z^{J_0} $, where $F_L$ is a would-be left-moving fermion number.
We do not try to make this precise.
Our normalization is motivated by the relation~(\ref{L-theta}).
}

For $\rho=AA$, the torus partition function with $U(1)$ fugacities is%See p.13 of Takuya's 19-12-27 paper note.
\begin{equation}\label{factorization_AA}
\begin{aligned}
&\quad \
  Z[(\mathcal{T}_\phi\times \mathcal{T}_\text{Arf})/\mathbb{Z}_2^\text{diag};R=2 (p/p'')^{1/2} ; \rho=AA]\\
  &= 
\frac{1}{|\eta(\tau)|^2} \sum_{m,\bar{m}\in\mathbb{Z}} 
q^{
\frac{1}{8} \left(
m\left( \frac{2}{R} +\frac{R}{2}\right) + \bar{m} \left( \frac{2}{R} -\frac{R}{2}\right)
\right)^2
}
z^{
\frac{1}{2k^{1/2}} \left(
m\left( \frac{2}{R} +\frac{R}{2}\right) + \bar{m} \left( \frac{2}{R} -\frac{R}{2}\right)
\right)
}
\\
&\qquad\times
\bar{q}^{
\frac{1}{8} \left(
\bar{m}\left( \frac{2}{R} +\frac{R}{2}\right) + m \left( \frac{2}{R} -\frac{R}{2}\right)
\right)^2
}
\bar{z}^{
\frac{1}{2k^{1/2}} \left(
\bar{m}\left( \frac{2}{R} +\frac{R}{2}\right) + m \left( \frac{2}{R} -\frac{R}{2}\right)
\right)
}
\\
&=
\frac{1}{|\eta(\tau)|^2} \sum_{m,\bar{m}\in\mathbb{Z}} 
q^{
\frac{1}{2k} (m p_1 + \bar{m} p_2)^2
}
z^{(m p_1 + \bar{m} p_2)/k}
\bar{q}^{
\frac{1}{2k} (m p_2 + \bar{m} p_1)^2
}
\bar{z}^{(\bar{m} p_1 + m p_2)/k}
\\
&=
\frac{1}{|\eta(\tau)|^2} \sum_{\lambda=0}^{k-1} \sum_{n,\bar{n}\in\mathbb{Z}} 
q^{
\frac{1}{2k} (k n +\lambda)^2
}
z^{ n +(\lambda/k)}
\bar{q}^{
\frac{1}{2k} (k \bar{n} +\omega \lambda)^2
}
\bar{z}^{ \bar{n} +\omega (\lambda/k)}
\\
&=
\sum_{\lambda=0}^{k-1} 
L^{(0,0)}_\lambda(\nu,\tau)  \overline{L^{(0,0)}_{\omega\lambda}(\nu, \tau)}
\,.
\end{aligned}
\end{equation}
Similarly for $AP$,
\begin{equation}\label{factorization_AP}
\begin{aligned}
&\quad \ 
  Z[(\mathcal{T}_\phi\times \mathcal{T}_\text{Arf})/\mathbb{Z}_2^\text{diag};R=2 (p/p'')^{1/2} ; \rho=AP]\\
&=
\frac{1}{|\eta(\tau)|^2} \sum_{m,\bar{m}\in\mathbb{Z}} 
(-1)^{m+\bar{m}}
q^{
\frac{1}{8} \left(
m\left( \frac{2}{R} +\frac{R}{2}\right) + \bar{m} \left( \frac{2}{R} -\frac{R}{2}\right)
\right)^2
}
z^{
\frac{1}{2k^{1/2}} \left(
m\left( \frac{2}{R} +\frac{R}{2}\right) + \bar{m} \left( \frac{2}{R} -\frac{R}{2}\right)
\right)
}
\\
&\qquad\times
\bar{q}^{
\frac{1}{8} \left(
\bar{m}\left( \frac{2}{R} +\frac{R}{2}\right) + m \left( \frac{2}{R} -\frac{R}{2}\right)
\right)^2
}
\bar{z}^{
\frac{1}{2k^{1/2}} \left(
\bar{m}\left( \frac{2}{R} +\frac{R}{2}\right) + m \left( \frac{2}{R} -\frac{R}{2}\right)
\right)
}
\\
&=
\sum_{\lambda=0}^{k-1} 
(-1)^{(1+\omega)\lambda}
L^{(0,1)}_\lambda (\nu,\tau)
\overline{L^{(0,1)}_{\omega \lambda} ( z, \tau)}
\,.
\end{aligned}
\end{equation}
For $\rho =PA$,
\begin{equation}\label{factorization_PA}
\begin{aligned}
&\quad \ 
 Z[(\mathcal{T}_\phi\times \mathcal{T}_\text{Arf})/\mathbb{Z}_2^\text{diag};R=2 (p/p'')^{1/2} ; \rho=PA]\\
&=
\frac{1}{|\eta(\tau)|^2} 
\sum_{m,\bar{m}\in\mathbb{Z}} 
q^{
\frac{1}{8} \left(
m\left( \frac{2}{R} +\frac{R}{2}\right) + \bar{m} \left( \frac{2}{R} -\frac{R}{2} \right) + \frac{2}{R} 
\right)^2
}
z^{\frac{1}{2 k^{1/2}} \left[m\left( \frac{2}{R} +\frac{R}{2}\right) + \bar{m} \left( \frac{2}{R} -\frac{R}{2}  \right)+ \frac{2}{R}\right]}
\\
&\qquad\times
\bar{q}^{
\frac{1}{8} \left(
\bar{m}\left( \frac{2}{R} +\frac{R}{2}\right) + m \left( \frac{2}{R} -\frac{R}{2}  \right)+ \frac{2}{R}
\right)^2
}
\bar{z}^{\frac{1}{2 k^{1/2}} \left[\bar{m}\left( \frac{2}{R} +\frac{R}{2}\right) + m \left( \frac{2}{R} -\frac{R}{2}  \right)+ \frac{2}{R}\right]}
\\
&=
\sum_{\lambda=0}^{k-1}
L^{(1,0)}_{\lambda+p''/2} (\nu,\tau)
\overline{L^{(1,0)}_{\omega \lambda+p''/2} (\nu,\tau)}
\,.
\end{aligned}
\end{equation}
Finally for $\rho=PP$,
\begin{equation}\label{factorization_PP}
\begin{aligned}
&\quad \ 
 Z[(\mathcal{T}_\phi\times \mathcal{T}_\text{Arf})/\mathbb{Z}_2^\text{diag};R=2 (p/p'')^{1/2} ; \rho=PP]\\
&=
 \frac{-1}{|\eta(\tau)|^2} \sum_{m,\bar{m}\in\mathbb{Z}} 
 (-1)^{m+\bar{m}}
q^{
\frac{1}{8} \left(
m\left( \frac{2}{R} +\frac{R}{2}\right) + \bar{m} \left( \frac{2}{R} -\frac{R}{2} \right) + \frac{2}{R} 
\right)^2
}
z^{\frac{1}{2 k^{1/2}} \left[m\left( \frac{2}{R} +\frac{R}{2}\right) + \bar{m} \left( \frac{2}{R} -\frac{R}{2}  \right)+ \frac{2}{R}\right]}
\\
&\qquad\times
\bar{q}^{
\frac{1}{8} \left(
\bar{m}\left( \frac{2}{R} +\frac{R}{2}\right) + m \left( \frac{2}{R} -\frac{R}{2}  \right)+ \frac{2}{R}
\right)^2
}
\bar{z}^{\frac{1}{2 k^{1/2}} \left[\bar{m}\left( \frac{2}{R} +\frac{R}{2}\right) + m \left( \frac{2}{R} -\frac{R}{2}  \right)+ \frac{2}{R}\right]}
\\
&=
\sum_{\lambda=0}^{k-1}(-1)^{(1+\omega)\lambda+1}
L^{(1,1)}_{\lambda+p''/2} (\nu,\tau)
\overline{L^{(1,1)}_{\omega \lambda+p''/2} (\nu,\tau)}
\,.
\end{aligned}
\end{equation}

For $k=1$ our conformal blocks are related to Jacobi's theta functions~$\vartheta_{ab}(\nu,\tau)$ and the Dedekind eta function as
\begin{equation}\label{L-theta}
L^{(S_1,S_2)}_{\lambda=S_1/2}(\nu,\tau)=\frac{ \vartheta_{S_1 S_2}(\nu,\tau)}{\eta(\tau)} \,.
\end{equation}

\subsection{Modular matrices}

We note the relation
\begin{equation}
K_{\lambda+1/2}(z,\tau) = q^{\frac{\lambda}{2k} + \frac{1}{8k}}K_\lambda(q^{1/2} z, \tau) \,.
\end{equation}
For $\lambda\in \frac{1}{2}\mathbb{Z}$, the function $K_\lambda(z,\tau)$ transforms as
\begin{equation}\label{K-T-trans}
K_\lambda(z,\tau+1) = e^{- \frac{\pi i}{12}} e^{\frac{\pi i }{k} \lambda^2} K_\lambda( (-1)^{2\lambda+1} z,\tau) \,,
\end{equation}
\begin{equation}\label{K-S-trans}
K_\lambda(e^{2\pi i\nu/\tau},-1/\tau) =e^{\pi i\frac{ \nu^2}{k\tau}} \frac{1}{k^{1/2}} \sum_{\mu=0}^{k-1}  e^{-2\pi i \frac{\mu\lambda}{k}} e^{-2\pi i \frac{\lambda}{k} \frac{\nu}{\tau}} e^{2\pi i  \frac{\mu}{k}\nu} K_\mu((-1)^{2\lambda} e^{2\pi i \nu},\tau) \,.
\end{equation}%\TO{See p.3 of 12-19 and p.5 of 12-27 paper notes.}
These can be used to show, for $ \lambda\in\mathbb{Z} + \frac{S_1}{2}$,  that
\begin{equation} \label{L-T-transform}
L_\lambda^{(S_1,S_2)}(\nu,\tau+1) = e^{-\frac{\pi i }{12}} e^{\frac{\pi i}{k}\lambda^2} L^{(S_1,S_1+S_2+1)}_\lambda(\nu,\tau) 
\,,
\end{equation}
\begin{equation} \label{L-S-transform}
L^{(S_1,S_2)}_\lambda(\nu/\tau,-1/\tau) = e^{\pi i \frac{\nu^2}{k\tau}}  \frac{1}{k^{1/2}} \sum_{\mu=S_2/2}^{k-1+S_2/2} e^{-2\pi i \frac{\lambda \mu}{k}} L_\mu^{(S_2,S_1)} (\nu,\tau)  
\,.
\end{equation}
In terms of the functions $K^{(S_1,S_2)}_\lambda(\nu,\tau,\rho):= e^{-2\pi i \rho/k} L^{(S_1,S_2)}_\lambda(\nu,\tau)$ that depend on an extra parameter $\rho\in\mathbb{C}$, we can write these transformations as%
\footnote{%
The following is a well-defined action of the group $SL(2,\mathbb{Z})$.
\begin{equation}
\begin{pmatrix}
a & b \\
c & d 
\end{pmatrix}
 : (\nu;\tau;\rho) \mapsto \left( \frac{\nu}{c\tau+d }; \frac{a\tau+b}{c\tau+d}; \rho+ \frac{c|\nu|^2}{2(c\tau+d)}\right) \,.
\end{equation}
}
\begin{align}
K^{(S_1,S_2)}_\lambda(\nu,\tau+1,\rho) &=  \sum_{T_1,T_2,\mu} \bm{T}_{(S_1,S_2;\lambda)\, (T_1,T_2;\mu)} K^{(T_1,T_2)}_\mu(\nu,\tau,\rho) \,,
\\
K^{(S_1,S_2)}_\lambda\left(\frac \nu \tau,-\frac1\tau,\rho+\frac{\nu^2}{2\tau}\right)&=  \sum_{T_1,T_2,\mu} \bm{S}_{(S_1,S_2;\lambda)\, (T_1,T_2;\mu)} K^{(T_1,T_2)}_\mu(\nu,\tau,\rho) \,,
\end{align}
with the matrices $\bm{S}$ and $\bm{T}$ given as%
\footnote{%
The modular matrices in~\cite{Belov:2005ze} were corrected in~\cite{Stirling:2008bq}, and are related to ours by conjugation.
}
\begin{align} \label{S-matrix-k-odd}
\bm{S}_{(S_1,S_2;\lambda)\, (T_1,T_2;\mu)} &= 
\frac{1}{\sqrt k} \delta_{S_1 T_2}\delta_{S_2 T_1} e^{-\frac{2\pi i }{k} \lambda \mu}  \,,
\\
\label{T-matrix-k-odd}
\bm{T}_{(S_1,S_2;\lambda)\, (T_1,T_2;\mu)} &= 
\delta_{S_1 T_1} \delta^\text{mod 2}_{S_1+S_2+1,T_2}  \delta_{\lambda \mu}e^{-\frac{\pi i} {12}} e^{\frac{\pi i }{k} \lambda^2}  \,.
\end{align}
Here $S_i,T_i \in \{0,1\}$ and 
we have  $\lambda \in \{0,1,\ldots,k-1\} + S_1/2$, $\mu \in \{0,1,\ldots,k-1\} + T_1/2$.
The matrices $\bm{S}$ and $\bm{T}$ respectively represent the generators of $SL(2,\mathbb{Z})$%
\footnote{%
The matrices satisfy the defining relations,~$\bm{S}^4=1$ and $(\bm{ST})^3=\bm{S}^2$ of $SL(2,\mathbb{Z})$.
}
\begin{equation}
{\rm S} = \begin{pmatrix}
0 & -1 \\
1 & 0
\end{pmatrix}
\quad
\text{ and } 
\quad
{\rm T} = \begin{pmatrix}
1 & 1 \\
0 & 1
\end{pmatrix} \,.
\end{equation}

\section{$U(1)_{k=\text{odd}}$ Chern-Simons theory on $L(a,\pm 1)$}\label{sec:CS}

As an application of~(\ref{S-matrix-k-odd}) and~(\ref{T-matrix-k-odd}), we compute the partition function of the~$U(1)_k$ Chern-Simons theory, with $k$ odd, on the lens space~$L(a,\pm 1)$.

\subsection{Gluing matrix for $L(a,b)$}\label{sec:Lab}

We begin, rather pedantically, by reviewing how the general lens space~$L(a,b)$ ($a>0$, $a$ and $b$ relatively prime) is obtained by gluing two copies of solid torus.
Let us view the three-sphere $S^3$ as a subset $S^3 = \big\{(z_1,z_2)\in\mathbb{C}^2 \big| |z_1|^2 + |z_2|^2=1 \big\}$ of $\mathbb{C}^2$.
We take two integers $a\geq 1$ and $b$ that are relatively prime.
There exists a non-unique pair of integers~$a'$ and $b'$ such that $bb'-aa'=1$.
The lens space $L(a,b)$ is defined as a $\mathbb{Z}_a$ quotient
\begin{equation}
L(a,b) :=  S^3/\mathbb{Z}_a \,,
\end{equation}
where the $\mathbb{Z}_a$-action depends on $b$ and is specified by the action of the generator
\begin{equation}\label{Z_a action}
(z_1,z_2) \mapsto (e^{\frac{2\pi i }{a}} z_1 , e^{\frac{2\pi i  b}{a}} z_2)  \,.
\end{equation}
The lens space $L(a,b)$ can be obtained by gluing two copies of solid torus $D^2\times S^1$:
\begin{equation}\label{Lab-glueing}
L(a,b) \simeq  \Big((D^2\times S^1)_1 \cup (D^2\times S^1)_2 \Big)\Big/\sim \,,
\end{equation}
where the boundaries of the first copy 
\begin{equation}\label{solid-torus-1}
(D^2\times S^1)_1 := \big\{ (r e^{i\phi_1}, e^{i\phi_2}) \in \mathbb{C}^2 \, \big |\, 0\leq r\leq 1 \,, \ \phi_1,\phi_2\in [0,2\pi)\big\} 
\end{equation}
and the second copy
\begin{equation}\label{solid-torus-2}
(D^2\times S^1)_2 := \big\{ (  \tilde r e^{i \tilde \phi_1}, e^{i \tilde \phi_2}) \in \mathbb{C}^2\, \big| \,0\leq \tilde r\leq 1 \,, \ \tilde \phi_1,\tilde \phi_2\in [0,2\pi)\big\} 
\end{equation}
at $r=\tilde{r}=1$
are identified via the relation%
\footnote{%
In terms of the variables $(\psi_1,\psi_2)$ such that $z_1 = \cos\frac{\theta}{2} e^{i\psi_1}$, $z_2 = \sin\frac{\theta}{2} e^{i\psi_2}$, we have the relations
\begin{equation}
\begin{pmatrix}
\psi_1 \\
\psi_2
\end{pmatrix}
=
\begin{pmatrix}
1 &  b'/a \\
0 &  1/a
\end{pmatrix}
\begin{pmatrix}
\phi_1 \\
\phi_2
\end{pmatrix}
=
\begin{pmatrix}
0 &  1/a \\
-1 &  b/a
\end{pmatrix}
\begin{pmatrix}
\tilde \phi_1 \\
\tilde \phi_2
\end{pmatrix} 
\,.
\end{equation}
}
\begin{equation}\label{eq:gluing-U}
\begin{pmatrix}
\tilde \phi_1 \\
\tilde \phi_2
\end{pmatrix}
=
U
\begin{pmatrix}
\phi_1 \\
\phi_2
\end{pmatrix}
\text{ mod } 2\pi
\,,
\quad
U=
\begin{pmatrix}
b & a' \\
a & b'
\end{pmatrix} \in SL(2,\mathbb{Z})\,.
\end{equation}
We note that the shift $(b,a') \rightarrow (b+ ja,a'+jb')$ corresponds to multiplying $U$ by $T^j$ from the left.
Similarly the shift $(a',b') \mapsto (a' + jb, b' +j a)$ corresponds to the multiplication by $T^j$ from the right.

The transformation
\begin{equation}\label{phi-transform}
\begin{pmatrix}
\phi_1 \\
\phi_2
\end{pmatrix}
\mapsto
M
\begin{pmatrix}
\phi_1 \\
\phi_2
\end{pmatrix}
\,,
\qquad
M=
\begin{pmatrix}
\alpha & \beta \\
\gamma & \delta
\end{pmatrix} 
\in SL(2,\mathbb{Z})
\end{equation}
on the angular coordinates of the two-dimensional torus%
\footnote{%
The meridian $\partial D^2\times pt$ and the longitude $pt \times S^1$ are parametrized by $\phi_1$ and $\phi_2$, respectively.
This implies that our matrix convention in~(\ref{phi-transform}) is consistent with those of~\cite{Freed:1991wd,MR1175494,2002math......9403H}.
In addition, the transformation~(\ref{tau-transform}) of $\tau$ is consistent with~\cite{Witten:1988hf} and the CFT literature.
} 
induces the transformation
\begin{equation}\label{tau-transform}
\tau \mapsto \frac{\alpha\tau +\beta}{\gamma\tau+\delta}
\end{equation}
of the modulus $\tau$ defined as the ratio $\tau=\omega_2/\omega_1$ of two periods $\omega_1,\omega_2\in \mathbb{C}\backslash\{0\}$ such that ${\rm Im}(\omega_2/\omega_1)>0$ if we adopt the convention where $(\phi_1,\phi_2)$ corresponds to a point $(\phi_2\omega_1-\phi_1\omega_2)/2\pi$ on the complex plane~$\mathbb{C}$.

For a bosonic TQFT such as the $SU(2)_k$ Chern-Simons theory considered in~\cite{Witten:1988hf}, the partition function on $L(a,b)$ is computed as follows.
Let $\mathcal{F}_i(\tau)$ be the torus conformal blocks (characters) of the edge state CFT.
Under the transformation~(\ref{tau-transform}), the blocks behave as
\begin{equation}
\mathcal{F}_i
\left( \frac{\alpha\tau +\beta}{\gamma\tau+\delta}\right)
= \sum_j \bm{M}_{ij} \mathcal{F}_j(\tau) \,,
\end{equation}
where the matrix $\bm{M}=(\bm{M}_{ij})$ represents $M$ on the space spanned by the conformal blocks.
In~\cite{Witten:1988hf} 
 a formula was given in terms of the matrix $U$ for the $L(a,b)$ partition function of the Chern-Simons theory.
It reads
\begin{equation} \label{lens-PF-modular even}
Z[L(a,b)] = {\bm U}_{00} \,,
\end{equation}
where
$0$ denotes the identity state, and~$\bm{U}$ represents the gluing matrix~$U$ in~(\ref{eq:gluing-U}).
In this note we choose to ignore framing dependence~\cite{Witten:1988hf}.

\subsection{Spin structures in 2d and 3d}

For a spin TQFT, the partition function depends on the choice of a spin structure.

An orientable manifold admits a spin structure if and only if its second Stiefel-Whitney class $w_2$ vanishes.
Any orientable 3-manifold admits a spin structure because its tangent bundle is trivial (and hence $w_2=0$).
The cohomology~$H^1(X,\mathbb{Z}_2)$ of any manifold $X$ acts on the spin structures on $X$ transitively and freely, and thus classifies them.
(We stated this fact in terms of $\mathbb{Z}_2$ gauge fields in Section~\ref{sec:Arf}.) % See p.1058 of "Cohomology algebra of orbit spaces of free involutions on lens spaces" by M. Singh.
The cohomology~$H^1(L(a,b),\mathbb{Z}_2)$ is 0 for $a$ odd and $\mathbb{Z}_2$ for $a$ even.

Thus $L(a,b)$ admits a single spin structure for $a$ odd and two spin structures for $a$ even.
This can also be seen in terms of spin structures on the 2d torus as follows.

The lens space~$L(a,b)$ is obtained by gluing two copies of the solid tori as in~(\ref{Lab-glueing}).
Their boundaries $(T^2)_1:=\partial(D^2\times S^1)_1$ and $(T^2)_2:=\partial(D^2\times S^1)_2$ carry 2d spin structures.
A spin structure on~$T^2$ is specified by periodic ($P \leftrightarrow 1$) or anti-periodic ($A \leftrightarrow 0$) boundary conditions (for an auxiliary spinor) along the two circles.
The 2d spin structures on the two tori that are compatible with gluing give rise to a 3d spin structure~$\rho_\text{3d}$ on~$L(a,b)$.
Let us study the boundary conditions on the auxiliary spinor
\begin{equation}
\Psi(\phi_1,\phi_2) = \tilde{\Psi}(\tilde{\phi}_1, \tilde{\phi}_2)
\end{equation}
with coordinates related via~(\ref{eq:gluing-U}).
We have
\begin{equation}
\begin{aligned}
\Psi(\phi_1,\phi_2) &=  (-1)^{1+S_1} \Psi(\phi_1+2\pi,\phi_2) = (-1)^{1+S_2} \Psi(\phi_1,\phi_2+2\pi)\,,  \\
\tilde\Psi(\tilde\phi_1,\tilde\phi_2) &=  (-1)^{1+T_1} \Psi(\tilde\phi_1+2\pi,\tilde\phi_2) = (-1)^{1+T_2} \Psi(\tilde\phi_1,\tilde\phi_2+2\pi) \,.
\end{aligned}
\end{equation}
We see that the spin structure $\rho=(S_1,S_2)\cdot \rho_0$ on $(T^2)_1$ and another $\sigma=(T_1,T_2)\cdot \rho_0$ on $(T^2)_2$ are related as
\begin{equation} \label{eq:sstilde-relations}
\begin{array}{ll}
\vspace{-3mm}
S_1 =&  1+ (1- T_1) b + (1- T_2) a  \\
\\
S_2  =&  1+ (1- T_1) a' + (1- T_2) b'   
\end{array}
\quad
 \text{ mod } 2 \,.
\end{equation}

It is well known that the boundary condition along the boundary of a disk is necessarily anti-periodic ($A\leftrightarrow0$) corresponding to the NSNS sector~\cite{Polchinski:1998rr}.
Thus we necessarily have $S_1=T_1=0$.
For a given $U$ in~(\ref{eq:gluing-U}) with $bb'-aa'=1$, the equations~(\ref{eq:sstilde-relations}) then admit two solutions for $S_2$ and $T_2$ if $a$ is even, and a unique solution if $a$ odd:
\begin{equation} \label{possible-2d-spin-structures}
\left.
\begin{array}{lllllll}
a =0  &\Rightarrow& b=b'=1  &\Rightarrow  & S_2=0 \text{ or } 1\,, &   T_2= a'+ S_2
\\
a =1  &\Rightarrow& a'=bb'+1  &\Rightarrow  & S_2= b'\,, &  T_2 = b  
\end{array}
\right\}
\text{ all mod 2.}
\end{equation}
Thus the lens space $L(a,b)$ admits two spin structures for $a$ even and a single spin structure for $a$ odd, as expected.

We expect that for a general spin TQFT, the edge state CFT with a chiral algebra symmetry is a spin CFT whose conformal blocks depend on the spin structure~$\rho$ on the surface, as in~(\ref{sec:character}).
Let~$\mathcal{F}_{\rho,i}(\tau)$ denote the torus conformal blocks that depend on the spin structure $\rho$, in addition to  the representation $i$ of the algebra and the modulus $\tau$.
The modular matrices are introduced by
\begin{equation}
\mathcal{F}_{\rho,i}(-1/\tau) = \sum_\sigma \sum_j \bm{S}_{(\rho,i)( \sigma, j)}  \mathcal{F}_{\sigma,j}(\tau) \,,
\quad
\mathcal{F}_{\rho,i}(\tau+1) = \sum_\sigma \sum_j \bm{T}_{(\rho,i)( \sigma, j)}  \mathcal{F}_{\sigma,j}(\tau) \,,
\end{equation}
\begin{equation}
\mathcal{F}_{\rho,i}\left(\frac{b\tau + a'}{a\tau+ b'} \right) = \sum_\sigma \sum_j \bm{U}_{(\rho,i)( \sigma, j)}  \mathcal{F}_{\sigma,j}(\tau) \,.
\end{equation}
The matrix $\bm{U}$ represents the gluing matrix~$U$ in~(\ref{eq:gluing-U}) for the lens space.
We propose that the lens space partition function of the spin TQFT for a given 3d spin structure $\rho_\text{3d}$ is given by
\begin{equation} \label{lens-PF-modular odd}
Z[L(a,b);  \rho_\text{3d}] =  {\bm U}_{(\rho,0)(\sigma,0)} \,,
\end{equation}
where $0$ denotes the representation that contains the ground state in the given sector, and the pair $(\rho,\sigma)$ of 2d spin structures corresponds to the 3d spin structure~$\rho_\text{3d}$.

\subsection{Partition function on lens space~$L(a,\pm1)$}

For the lens space~$L(a,\epsilon=\pm1)$, the gluing matrix is
\begin{equation}\label{U-Laep}
U = 
\begin{pmatrix}
\epsilon & 0 \\
a & \epsilon
\end{pmatrix} 
= S^{\epsilon} T^{-\epsilon a} S^{-1}
\,.
\end{equation}
Using the formulas~(\ref{S-matrix-k-odd}) and~(\ref{T-matrix-k-odd}) we obtain, for $\bm{U}=\bm{S}^\epsilon \bm{T}^{-\epsilon a} \bm{S}^{-1}$,
\begin{equation}
\bm{U}_{(S_1,S_2;\lambda) \, (T_1,T_2;\mu)} =e^{\frac{\epsilon a}{12}\pi i} \delta^\text{mod 2}_{S_1+a(S_2+1), T_1}\delta_{S_2 T_2} \frac{1}{k} \sum_{\nu = S_2/2}^{k-1+S_2/2} e^{\frac{2\pi i }{k} \nu (-\epsilon \lambda + \mu)}  e^{-\frac{\epsilon a}{k} \pi i \nu^2} \,.
\end{equation}
Here $S_i,T_i \in \{0,1\}$ and 
we have  $\lambda \in \{0,1,\ldots,k-1\} + S_1/2$, $\mu \in \{0,1,\ldots,k-1\} + T_1/2$.

Let us recall from~(\ref{possible-2d-spin-structures}) that the possible spin structures on the 2d torus depend on~$a$.
For both $a$ even and odd, there exists a 3d spin structure corresponding to the 2d spin structures~$\rho=\sigma=(0,1)\cdot\rho_0=AP$
 for which we obtain via the reciprocity formula for the Gauss sum:%
\footnote{%
%In this footnote, the definitions of various symbols do not coincide with those in the main text.
Let $\alpha,\beta,\gamma$ be integers with $\alpha\gamma\neq0$ and $\alpha \gamma+\beta$ even.  Then
(see for example Section~1.2 of~\cite{MR1625181})
\begin{equation} \label{reciprocity-formula}
\sum_{n=0}^{|\gamma|-1} e^{\pi i \frac{ \alpha n^2+\beta n}{\gamma}} =  \left|\frac{\gamma}{\alpha}\right|^{1/2} e^{\frac{\pi i}{4} (\text{sgn}(\alpha\gamma)- \frac{\beta^2}{\alpha\gamma})}
\sum_{n=0}^{|\alpha|-1} e^{-\pi i \frac{ \gamma n^2 + \beta n}{\alpha}}  \,.
\end{equation}
}
\begin{equation}\label{Z-APAP}
Z = \bm{U}_{(0,1;0)\,(0,1;0)}= \frac{e^{ \frac{\epsilon a}{12}\pi i} e^{- \frac{\pi i }{4} \epsilon}}{\sqrt{ka}} \sum_{h=0}^{a-1} e^{\epsilon \pi i \frac{k}{a} h^2} e^{-\pi i h} 
= \frac{e^{ \frac{\epsilon a}{12}\pi i} e^{- \frac{\pi i }{4} \epsilon}}{\sqrt{ka}} \sum_{h=0}^{a-1} e^{\epsilon \pi i \frac{k}{a}(1-a) h^2} \,.
\end{equation}
We interpret $h$ as a label for the gauge bundles that we sum over in the path integral.%
\footnote{%
The bundles can be specified either by the $U(1)$ holonomy around the generator of~$\pi_1(L(a,\pm 1))=\mathbb{Z}_a$, or by the first Chern class~$c_1(L)\in H^2(L(a,\pm1),\mathbb{Z})$.}
For $a$ even, there is another 3d spin structure corresponding to $\rho=\rho'=(0,0)\cdot\rho_0=AA$,
which gives%
\footnote{%
The difference between $e^{\epsilon \pi i \frac{k}{a}(1-a) h^2}$ in~(\ref{Z-APAP}) and $e^{\epsilon \pi i \frac{k}{a} h^2}$ in~(\ref{Z-AAAA}) is identical to the difference noted in footnote 2 of~\cite{Imamura:2013qxa}.}
\begin{equation}\label{Z-AAAA}
Z = \bm{U}_{(0,0;0)\,(0,0;0)}= \frac{e^{ \frac{\epsilon a}{12}\pi i} e^{- \frac{\pi i }{4} \epsilon}}{\sqrt{ka}} \sum_{h=0}^{a-1} e^{\epsilon \pi i \frac{k}{a} h^2}  \,.
\end{equation}%\TO{Checked in Compare-U(1)k-odd-partition-functions.nb on 03/23.}
We see that $\epsilon=1$ and $\epsilon=-1$ are related by complex conjugation as expected;  a change of orientation should replace the partition function by its complex conjugate~\cite{MR1001453}.

We note that the sums in~(\ref{Z-APAP}) and~(\ref{Z-AAAA}) involve the so-called quadratic refinements of the linking pairing on~$H^2(L(a,\pm1),\mathbb{Z})=\mathbb{Z}_a$.
Let us consider the bilinear paring $B: \mathbb{Z}\times\mathbb{Z}\rightarrow\mathbb{Z}$ defined by $B(x,y)= \epsilon a xy$.
Correspondingly~\cite{MR324709} we have the bilinear pairing $L: \mathbb{Z}_a \times \mathbb{Z}_a \rightarrow \mathbb{Q}/\mathbb{Z}$ given by $L(h,h')= h h'/\epsilon a$ mod $\mathbb{Z}$.
For $v\in\{0,1\}$, let us try to define the map $\psi_v : \mathbb{Z}_a \rightarrow \mathbb{Q}/\mathbb{Z}$ defined by
\begin{equation}
\psi_v(h):= \frac{1}{2} h \left(\frac{1}{\epsilon a} h + v\right)  \text{ mod } \mathbb{Z}\,. 
\end{equation}
When $a$ is odd, $\psi_0$ is well-defined and is the unique quadratic refinement of $L$.
When $a$ is even, both $\psi_0$ and $\psi_1$ are well-defined and are the only possible quadratic refinements of~$L$.
The integer $v$ is called a Wu class for $B$~in~\cite{MR324709}.
The exponentials in (\ref{Z-APAP}) and~(\ref{Z-AAAA}) involve $\psi_1$ and $\psi_0$, respectively.

\section{Discussion}

\subsection{Comments on the relation between conformal blocks and Wilson line operators}\label{sec:block-line}

The spin structure-dependent conformal blocks we introduced in~(\ref{def:conformal-blocks}) are similar to the wave functions obtained by the canonical quantization of Chern-Simons theory in~\cite{Belov:2005ze}.
They should be interpreted as providing basis states for the spin structure-dependent Hilbert spaces for the 3d spin TQFT.
For a 3d non-spin Chern-Simons theory, a torus conformal block of the corresponding Wess-Zumino-Witten model is identified with a state obtained by the path integral on a solid torus with a Wilson line operator inserted along a non-contractible cycle~\cite{Witten:1988hf}.%
\footnote{%
For $k$ even the Wilson operator $\exp \mu \oint A$ is identified with $\exp (\mu+k)\oint A$.
The corresponding conformal blocks~$K_\mu$ defined in~(\ref{def:K-lambda}) also satisfy $K_{\mu+k}=K_{\mu}$ for $\mu\in\mathbb{Z}$.
}

We observe that such a correspondence does not hold for the $U(1)$ spin Chern-Simons theory and for the spin structure-dependent basis states on a torus, at least when the following classification of Wilson lines is assumed.
The Wilson operator $W_\mu:=\exp \mu \oint A$ has topological spin~$\theta_\mu:=\exp \frac{\pi i}{k}\mu^2$.
Then for $\mu$ an integer,~$W_{\mu+k}$ has topological spin~$\theta_{\mu+k}=\theta_\mu (-1)^k$, which differs from $\theta_\mu$ by a sign for $k$ odd.
For this reason one usually classifies Wilson operators by $\mu\in\mathbb{Z}$ with the identification~$\mu\sim\mu+2k$.
See for example~\cite{Seiberg:2016rsg}.
The operator~$W_k$ has topological spin $-1$ and represents a transparent fermion line.
(For~$k=1$,~$W_1$ corresponds to the 2d Dirac fermion, which is $e^{i\phi_L}$ upon bosonization.)
The spin structure-dependent conformal blocks in~(\ref{def:conformal-blocks}) are instead labeled by $\lambda\in \frac{1}{2}\mathbb{Z}$ with the restriction~$0\leq \lambda<k$.
(For $k=1$, the Dirac fermion, which is in the NS sector and is nothing but the fermionic current for the extended chiral superalgebra~\cite{Dijkgraaf:1989pz}, contributes to the conformal blocks~$L^{(S_1=0,S_2)}_{\lambda=0}=\vartheta_{0S_2}/\eta$ with $S_2=0,1$.)

The structure we found, where the identity and the transparent fermion are grouped into a single object, is also seen in the category-based approach~\cite{Aasen:2017ubm} (Table~2.4.1) to fermionic anyon condensation.%
\footnote{%
The fermion line $W_k$ of $U(1)_k$ comes from the fermion Wilson line $W^{(4k)}_{2k}$ of charge $2k$ in the bosonic parent theory $U(1)_{4k}$.
In the language of~\cite{Gaiotto:2015zta}, $W^{(4k)}_{2k}$ combines with the fermion line of an invertible spin TQFT called~$K_3$ to form a bosonic line that generates a non-anomalous $\mathbb{Z}_2$ one-form symmetry, gauging which leads to  $U(1)_k$.
Such a gauging is equivalent to condensing the boson line obtained from the two fermion lines.
}
We emphasize that spin structure-dependent conformal blocks are the basis relevant for computing 3d partition functions and topological invariants.

Instead of labeling conformal blocks by the temporal spin structure~$S_2$, one may alternatively consider the characters for (2d chiral) states with definite fermion numbers~\cite{Kapustin:2017jrc,Hsieh:2020uwb}.
Such a basis of 3d states would correspond to the insertion of a Wilson line in the solid torus.
(For~$k=1$ the identity and the Dirac fermion have different fermion numbers and are separated in the new basis.)
Still, to account for the conformal blocks with $S_1=1$ (Ramond boundary condition in the spatial direction), one would need to introduce Wilson lines whose charges are half odd integers and which are not gauge invariant in the usual sense.
The parent theory $U(1)_{4k}$ does have counterparts of these operators.
It would be interesting to understand how to make sense of such line operators, presumably with the help of spin structures, within the spin theory~$U(1)_k$.

\subsection{Other issues and future directions}\label{sec:other}

We focused in this paper on one of the simplest classes of spin TQFTs and the corresponding 2d CFTs.
It will be interesting to apply the analysis of this paper to more general 2d CFTs such as WZW models and compute the modular matrices and the partition functions of the corresponding spin TQFTs.
It also seems worthwhile to revisit the general $U(1)^N$ spin Chern-Simons theory~\cite{Belov:2005ze,2016PhRvB..94o5113L} and compare it with the boson CFT with the target space~$T^N$ modified by the Arf invariant.

For $k$ odd, the relation between the $U(1)_k$ and $U(1)_{4k}$ Chern-Simons theories mentioned in footnote~\ref{footnote-U14k} is an example of ``fermionic anyon condensation"~\cite{Gaiotto:2015zta,2017JMP....58d1704B,Aasen:2017ubm}.
It seems natural to combine our approach, base on the 2d Arf invariant, with the study of fermionic anyon condensation based on the study of appropriate  categories.
This will involve generalizing earlier results on bosonic anyon condensation (see for example \cite{2014NuPhB.886..436K,2014PhRvB..90s5130E,Shen:2019wop}) and incorporate results on fermionic (or super) fusion categories (see for example \cite{Gu:2010na,Aasen:2017ubm,2016arXiv160603466U}).

Our factorization results~(\ref{factorization_AA})-(\ref{factorization_PP}) should admit an interpretation in terms of a gapped domain wall in the spin $U(1)$ Chern-Simons theory, generalizing the results of~\cite{Kapustin:2010if} in the non-spin case.
Another worthwhile future direction would be the computation of the partition functions on more general lens spaces~\cite{Lab}.

As we mentioned at the end of Section~\ref{sec:Lab}, we ignored in this paper the dependence of the 3d partition function on the framing of the manifold~\cite{Witten:1988hf}.
The choice of $U$ in~(\ref{U-Laep}) for the lens space $L(a,\pm 1)$ implicitly specifies the framing.
In the non-spin case the framing dependence was studied in detail, for example, in~\cite{Freed:1991wd,MR1175494}.
It would be interesting to see whether new structures appear in the spin case.

\section*{Acknowledgements}
T.O. thanks S.~Monnier for helpful correspondence.
The works of T.O. and S.Y. are supported in part by JSPS KAKENHI Grant Numbers JP16K05312 and JP19K03847, respectively.

\appendix

\section{$k$ even}\label{app:even}

Recall that $R=2 (p/p'')^{1/2}$ and $k=pp''$.
Let us consider the case $p''=2p'$ even (hence $k$ even).
Since $p$ and $p'$ are relatively prime there exist integers $(r_0,s_0)$ such that $pr_0-p's_0=1$.
We set $\omega_0 =pr_0 +p's_0$.
Then we can expand the original torus partition function~(\ref{torus-pf-phi}) as
\begin{equation}
Z[\mathcal{T}_\phi] = \sum_{\lambda=0}^{k-1} K_\lambda(\tau) \overline{K_{\omega_0\lambda}( \tau)} \,,
\end{equation}
where
\begin{equation} \label{def:K-lambda}
K_\lambda(\tau):=\frac{1}{\eta(\tau)} \sum_{n\in\mathbb{Z}} q^{(kn+\lambda)^2/2k} \,.
\end{equation}
See Exercise~10.21 of~\cite{DiFrancesco:1997nk}.
We have
\begin{equation}
K_\lambda(-1/\tau) = \sum_{\mu=0}^{k-1} \bm{S}_{\lambda\mu} K_\mu(\tau) \,,
\quad
K_\lambda(\tau+1) = \sum_{\mu=0}^{k-1} \bm{T}_{\lambda\mu} K_\mu(\tau) \,,
\end{equation}
where
\begin{equation}
\bm{S}_{\lambda\mu} = \frac{1}{\sqrt k} e^{ -\frac{2\pi i}{k}\lambda\mu} \,,
\quad
\bm{T}_{\lambda\mu} :=
  e^{\pi i \frac{ \lambda^2}{k}} e^{-\frac{\pi i}{12}} \delta_{\lambda\mu} \,.
\end{equation}
See~(\ref{K-T-trans}) and~(\ref{K-S-trans}).
Using the reciprocity formula~(\ref{reciprocity-formula}) one can check that these give rise to a genuine (rather than projective) representation of $SL(2,\mathbb{Z})$.

For $L(a,\epsilon=\pm 1)$ the partition function is computed via~(\ref{lens-PF-modular even}) and~(\ref{reciprocity-formula}) as
\begin{equation}
Z= (\bm{S}^\epsilon\bm{T}^{-\epsilon a}\bm{S}^{-1})_{00} 
=\frac{e^{  \frac{ \epsilon a}{12} \pi i}}{k}\sum_{\nu =0}^{k-1} e^{-\epsilon \frac{\pi i}{k} a \nu^2} 
=\frac{e^{\epsilon \frac{\pi i}{12}a}e^{-  \frac{\pi i}{4}\epsilon }}{\sqrt{k a}}
\sum_{h=0}^{a -1} e^{ \epsilon \frac{\pi i}{a} k h^2} 
\,.
\end{equation}
This is the same expression as~(\ref{Z-AAAA}).
There is no dependence on the 3d spin structure.

\section{Quadratic refinements and their Arf invariants}\label{app:quad-Arf}

Let $K$ be a finite abelian group, and $L:K\times K\rightarrow \mathbb{Q}/\mathbb{Z}$ a symmetric bilinear pairing.
Here $\mathbb{Q}$ is the additive group of rational numbers.
We assume that $L$ is non-degenerate, meaning that if $L(x,y)=0$ for all $x\in K$ then $y=0$.

In this paper we define a {\it quadratic refinement} over $L$ to be a map $\psi: K\rightarrow \mathbb{R}/\mathbb{Z}$ such that%
\footnote{%
In some references such as~\cite{Belov:2005ze} the term ``quadratic refinement'' refers to a more general map $\psi': K\rightarrow \mathbb{R}/\mathbb{Z}$ satisfying $\psi'(x+y) -\psi'(x)-\psi'(y) +\psi'(0) = L(x,y)$ for $x, y\in K$.
%the definition of a quadratic refinement replaces~(\ref{brumfiel-morgan-1}) by $\psi(x+y) -\psi(x)-\psi(y) +\psi(0) = L(x,y)$ for $x, y\in K$ and drop the condition~(\ref{brumfiel-morgan-2}).
%Given such a $psi$, the function $\psi_\text{new}(x):=\psi(x)-\psi(0)$ satisfies (\ref{brumfiel-morgan-1}) and (\ref{brumfiel-morgan-2})?
A quadratic refinement in the sense of this paper is called a quadratic function over $L$ in~\cite{MR324709}.
%In the terminology used in \href{https://ncatlab.org/nlab/show/quadratic+refinement}{nLab}, our $\psi$ is the quadratic refinement of $L$ by a quadratic form.
Our terminology coincides with that of~\cite{MR1968575}.
We note that an identity similar to~(\ref{brumfiel-morgan-1}) is satisfied by the WZW action~\cite{Polyakov:1984et,Moore-TASI}.
}
\begin{equation} \label{brumfiel-morgan-1}
\psi(x+y) -\psi(x)-\psi(y) = L(x,y) \quad \text{ for }  x, y\in K
\end{equation}
and
\begin{equation}\label{brumfiel-morgan-2}
\psi(nx)=n^2\psi(x)\,, \quad n\in\mathbb{Z} \quad x \in K \,.
\end{equation}
Given (\ref{brumfiel-morgan-1}), the condition (\ref{brumfiel-morgan-2}) is equivalent to $2\psi(x)=L(x,x)$.

Following~\cite{MR324709}, we define the Arf invariant ${\rm Arf}(\psi)$ of a quadratic refinement $\psi$ by
\begin{equation}\label{Arf-quad-def}
e^{2\pi i {\rm Arf}(\psi)}:= \sum_{x\in K} e^{2\pi i\psi(x)}.
\end{equation}
It is known that ${\rm Arf}(\psi)$ takes values in $\mathbb{Q}/\mathbb{Z}$.

The Arf invariant ${\rm Arf}[\rho]$ for 2d spin structures~$\rho$, discussed in Section~\ref{sec:Arf}, fits the definition~(\ref{Arf-quad-def}) up to normalization.
Indeed the spin structures~$\rho$ on a closed surface $\Sigma$ are in one-to-one correspondence with the quadratic refinements $\psi_\rho$ of the pairing $L=(1/2)\phi$ on $K=H_1(\Sigma,\mathbb{Z}_2)$, where $\phi$ is the $\mathbb{Z}_2$-valued intersection form $(a,b)\mapsto \phi(a,b)= \# (a\cap b)$ mod $\mathbb{Z}_2$~\cite{MR588283}.
We have the relation
\begin{equation}
{\rm Arf}[\rho] = \frac{1}{2}{\rm Arf}(\psi_\rho)
\end{equation}
under the correspondence.

\bibliographystyle{utphys}
\bibliography{refs}

\end{document}